\documentstyle[12pt,aps,floats,epsf]{revtex}

\def\thru#1{\mathrel{\mathop{#1\!\!\!/}}}
\begin{document}
\tighten

\title{{HOW MUCH OF THE NUCLEON SPIN \\ IS CARRIED BY GLUE?}}

\author{Ian Balitsky$^{1,2,4}$ and Xiangdong Ji$^{3,4}$}
\vspace{ 0.1in}

\address{$^1$ Department of Physics \\
     Old Dominion University \\
     Norfolk, VA 23529}

\vspace{0.1in}
\address{$^2$ Jefferson Lab \\
      12000 Jefferson Ave.\\
      Newport News, VA 23606}
\vspace{0.1in}

\address{
$^3$ Department of Physics \\
University of Maryland \\
College Park, Maryland 20742
{~}}

\vspace{0.1in}
\address{ $^4$ Department of Physics \\
       Massachusetts Institute of Technology \\
       Cambridge, MA 02139}

\vspace{0.2in} 
\date{JLAB-THY-97-05~~~U.ofMD PP\#97-079 ~~~MIT-CTP-2607~~~January 1997}

\maketitle

\begin{abstract}
We estimate in the QCD sum rule approach the amount
of the nucleon spin carried by the gluon angluar 
momentum: the sum of the gluon spin and 
orbital angular momenta. The result indicates 
that gluons contribute at least one half of the 
nucleon spin at scale of 1 GeV$^2$. 

\end{abstract}
\pacs{xxxxxx}

\narrowtext

Ever since the publication of the EMC measurement 
on the fraction of the nucleon spin carried by the 
quark spin\cite{emc}, there has been a tremendous 
activity in the field of 
the spin structure of the nucleon\cite{review}. One of the 
central questions is how the
spin of the nucleon is distributed among its
constituents\cite{jaffemanohar}. After much debate, many agree now 
that a substential fraction of the 
nucleon spin comes from sources other than 
the quark spin, i.e., quark orbital and 
gluon angular momenta. Recently, several proposals have 
been made in the literature to measure the amount 
of the spin carried by the gluon helicity $\Delta G$\cite{spin96}.

In this Letter, we present a QCD sum rule calculation
\cite{shifman}
of the amount of the nucleon spin carried  
by gluons, or equivalently by quarks because, by definition,
their sum is 1/2. Our calculation is motivated 
by the possibility of measuring these quantities 
through deeply-virtual Compton scattering proposed by 
one of us \cite{ji1}. The method we use has been 
applied successfully to calculate a similar 
quantity---fractions of the nucleon momentum 
carried by quarks and gluons  
\cite{koles,belyaev}. Our result shows that the gluon 
angular momentum, the sum of gluon helicity and orbitial 
angular momentum, contributes at least 50\% of the nucleon 
spin, suggesting that the nucleon 
contains nontrivial gluon configurations carrying 
nonzero angular momentum.

The angular momentum operator in QCD can be written in an 
explicitly gauge invariant form \cite{ji1},
\begin{eqnarray}
     \vec{J}_{\rm QCD} &=& \int d^3 x~
        \Big[~ {1\over 2}\bar \psi \vec{\gamma}\gamma_5 \psi 
         + \psi^\dagger (\vec{x}\times (-i\vec{D}))\psi \nonumber \\
          & + & \vec{x}\times(\vec{E}\times\vec{B})\Big] \ . 
\end{eqnarray}
where flavor and color indices are implicit.  
The first term can be interpreted as the quark 
spin contribution, although its matrix element 
is actually the singlet axial charge. The second term, 
where the covariant derivative is ${\vec D} = \vec{\partial} 
+ ig\vec{A}$, is the canonical orbital 
angular momentum of quarks. The word ``canonical'' stems
from the {\it canonical} momentum for quarks in a background
gauge field. The last term is the total angular momentum 
of gluons, as is clear from the appearance of the 
Poynting vector. [In pure gauge theory without quarks, this term
generates the spin quantum numbers for glueballs]. 
According to the above expression, 
we can write down a gauge-invariant spin sum rule for the nucleon, 
\begin{equation}
      {1\over 2} = {1\over 2}
     \Delta \Sigma(\mu^2) + L_q(\mu^2) + J_g(\mu^2) \ , 
\end{equation}
where $\mu^2$ is a scale at which the operators are renormalized,
or more physically the nucleon wave function is probed. 
The first term is what has been measured in polarized
deep-inelastic scattering \cite{emc,altarelli}. The second and 
third terms represent quark orbital and
gluon contributions, respectively. We also introduce
the notion of the total quark contribution, 
$J_q = \Delta \Sigma/2 + L_q$, the sum of spin and orbital. 
By definition, both $J_q(\mu^2)$ and $J_g(\mu^2)$ 
are gauge-invariant if gauge-invariant regularization 
and renormalization schemes are used. In the light-like gauge $A^+=0$, 
$J_g(\mu^2)$ can be written as a sum of
the gluon helicity $\Delta G(\mu^2)$, measurable 
in polarized high-energy scattering\cite{spin96}, the gluon orbital 
angular momentum, as well as a term from quark-gluon 
interactions \cite{ji1}.
  
Before formulating the sum rule calculation, it is
instructive to review a derivation of Eq. (1).
The angular momentum operators of QCD are identified with    
the generators of the Lorentz group: $ J^{\mu\nu}$,
which in turn are defined from the angular momentum 
density $M^{\mu\nu\alpha}$ through,
\begin{equation}
       J^{\mu\nu} = \int d^3 \vec{x}~ M^{0\mu\nu}(\vec{x}) \ . 
\label{gene}
\end{equation}
The angular momentum density can be expressed in
terms of the symmetric, conserved energy-momentum 
tensor $T^{\alpha\beta}$,
\begin{equation}
      M^{\mu\nu\alpha} = T^{\mu\alpha} x^\nu
         - T^{\mu\nu}x^\alpha \ . 
\label{mden}
\end{equation}
The energy-momentum tensor of QCD can be written as a
sum of the quark and gluon parts,
\begin{equation}
      T^{\alpha\beta}  =
          T_q^{\alpha\beta} + T_g^{\alpha\beta} 
      = {1\over 4}\bar \psi \gamma^{(\alpha} 
        i\stackrel{\leftrightarrow}{D^{\beta)}} \psi
         + \left({1\over 4}g^{\alpha\beta}F^2 -F^{\alpha\mu}
          F^\beta_{~\mu}\right) \ ,  
\label{emt}
\end{equation}
where $(\alpha\beta)$ means symmetrization of the 
indices. It is then simple to see that the quark and 
gluon parts of the angular momentum operators in Eq. (1)
are derived from Eqs. (\ref{gene}) and (\ref{mden}) 
by substituting in the quark and gluon parts
of the energy-momentum tensor, respectively.          

According to the above, we can formulate the sum 
rule calculation of $J_g(\mu^2)$, or equivalently 
$J_q(\mu^2)$, in terms of the energy-momentum 
tensor $T^{\alpha\beta}_{q,g}$. 
Consider the following three-point correlation
function in the QCD vacuum, 
\begin{equation}
    W^{\mu\nu\alpha}_g(p)= \int d^4xd^4z ~\langle 0|
            T[\eta(x)\bar \eta(0)
           M^{\mu\nu\alpha}_g(z)]|0\rangle~ e^{ip\cdot x} \ , 
\end{equation}
where $M_g^{\mu\nu\alpha}$ is defined as in Eq. (\ref{mden}) 
with $T^{\alpha\beta}$ replaced by its gluonic part, and
$\eta(x)$ is the interpolating field for the nucleon, which we
choose to be \cite{ioffe},
\begin{equation}
    \eta(x) = \epsilon^{ijk}\left(u^{iT}C\gamma^\alpha u^j\right) 
     \gamma_5\gamma_\alpha d^k \ . 
\end{equation}
$W^{\mu\nu\alpha}$ contains
a nucleon double-pole contribution, with its residue 
proportional to $J_g(\mu^2)$, 
\begin{equation}
       W^{\mu\nu\alpha}_g = {J_g(\mu^2) \lambda_N^2\over 
         (p^2-m_N^2)^2} (2ip^\mu\gamma^\nu \thru p \gamma^\alpha) + ... \ , 
\end{equation}
where ellipses include nucleon double poles of
different Dirac structures, nucleon single poles, and
other dispersive contributions.
$\lambda_N$ is the nucleon decay constant corresponding to 
the interpolating current,
\begin{equation}
            \langle 0|\eta(0)|N(p)\rangle = \lambda_N U(p)\ . 
\end{equation}
In the following, we first calculate $W^{\mu\nu\alpha}$ in the 
deep-Euclidean region $-p^2>\!>\Lambda_{\rm QCD}^2$ using operator
product expansion (OPE), from which 
we attempt to extract the double-pole residue $J_g(\mu^2)$.

To ensure $J_q(\mu^2) +J_g(\mu^2) =1/2$ in the sum rule 
calculation, we use an implicit form of Ward identity, 
\begin{equation}
      T^{\alpha\beta} = \partial_\rho (T^{\rho\beta} x^\alpha)
           - x^\alpha \partial_\rho T^{\rho\beta} \ , 
\end{equation}
to we rewrite the three-point correlation function as,
\begin{equation}
         W^{\mu\nu\alpha}_g 
       = \int d^4x d^4z ~\langle 0|T\eta(x)\bar\eta(0)
         z^\mu \left(z^\alpha \partial_\rho 
         T^{\rho \nu}_g - z^\nu \partial_\rho 
       T^{\rho\alpha}_g\right)|0\rangle
     e^{ip\cdot x} \ . 
\end{equation}
From Eq. (\ref{emt}), we find,
\begin{equation}
    \partial_\rho T^{\rho\nu}_g = 
       - \bar \psi gF^{\nu\alpha}\gamma_\alpha\psi + ... \ , 
\end{equation}
where ellipses denote terms vanishing after using 
gluon's equations of motion. Thus, we arrive at a new 
form of the Green's function
\begin{equation}
     W^{\mu\nu\alpha}_g 
   =  \int d^4x d^4z 
    ~z^\mu z^\nu \langle 0|T \eta(x)\bar \eta(0)
  \hat {\cal O}^\alpha(z)|0\rangle e^{ipx}
    - (\nu\leftrightarrow\alpha) + ... \ , 
\end{equation}
where $\hat {\cal O}^\alpha(z) = \bar\psi gF^{\alpha\beta}
\gamma_\beta \psi(z)$. If one goes through a similar
procedure for a correlator with the quark part of the
energy momentum tensor, one finds that it can be reduced
to the same term with a negetive sign
plus a two-point nucleon correlation function with a 
double-pole residue 1/2. 

\begin{figure}[htb]
\mbox{
\epsfxsize=8cm
\epsfysize=3cm
\hspace{4cm}
\epsffile{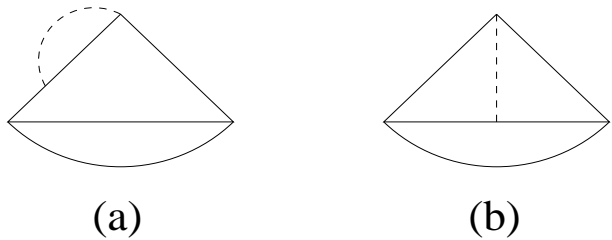}}
\vspace{0.3cm}
{\caption{\label{fig:fig1} Perturbative diagrams. 
Dashed line denotes gluon. 
(Permutations are not shown)}}
\end{figure}

The Green's function in the deep Euclidean space can be
calculated in OPE because of 
asymptotic freedom. The first term in such an expansion
is the usual perturbative contribution, which is infrared 
finite due to the finite external momentum $p^2$. There are two
perturbative diagrams as shown in Fig. 1.
We find the 
contribution from the first diagram  as,
\begin{equation}
    {\alpha_s\over \pi^5}\left({1\over 144}
    \ln^2{-p^2\over \mu^2} - {1\over 36}\ln{-p^2\over \mu^2}
    \right)p^2 \ , 
\end{equation}
where and henceforth we omit the structure factor 
$2ip^\mu\gamma^\nu\thru p \gamma^\alpha$. 
A calculation for the second diagram (``salboat") is rather tedious.
Since in the final result the (typical) contribution from 
the first diagram is small (less than 10\%), 
we discard this ``sailboat" contribution in the following 
study.
 
\begin{figure}[htb]
\mbox{
\epsfxsize=12cm
\epsfysize=3cm
\hspace{2cm}
\epsffile{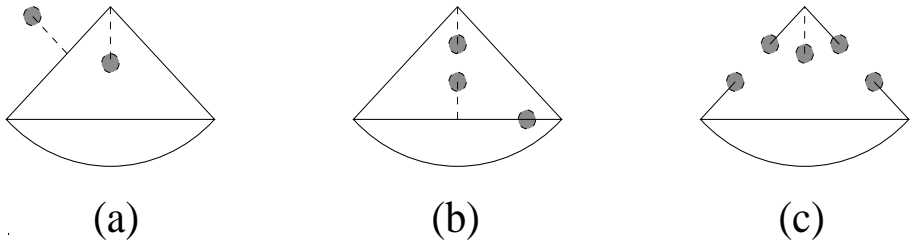}}
\vspace{0.3cm}
{\caption{\label{fig:fig2} Dimension-4 power corrections: local (a,b)
and bilocal (c). Shaded circles mark vacuum fields }}
\end{figure}

The next term in OPE comes
from dimension-four vacuum condensates. Diagrams from 
Fig. \ref{fig:fig2}a,b are found to contribute,
\begin{equation}
     -{1\over 144\pi^2p^2} \langle {\alpha_s\over \pi}F^2\rangle
       \left(\ln{-p^2\over \mu^2} +{7\over 6} + \ln{\mu^2
    \over -q^2} \right)\ , 
\label{dim4}
\end{equation}
where $q^2$ is an infrared regulator which represents
the momentum flow through the operator ${\cal O}^\alpha$. 
The infrared logarithm
arises from large separations of point 
$z$ from $0$ and $x$. To take into account the contribution
in this region properly, one must first expand the product 
of the interpolating current,
\begin{equation}
    T\eta(x)\bar \eta(0) = \sum_n C_n(x) \hat O_n\ , 
\end{equation}
(where $\hat O_n$ are a set of local operators) resulting in so-called 
bilocal power corrections \cite{bali3} (see Fig. \ref{fig:fig2}c). 
The relavant local
operator in this case is a dimension-five one,
\begin{eqnarray}
        \hat O_5^{\lambda\rho\rho'} & =&  
          2\bar u gF^{\lambda[\rho}\gamma^{\rho']}u
        -2i\partial^{[\rho}(\bar u \stackrel{\leftrightarrow}{D}^\lambda
           \gamma^{\rho']}u)
        + \bar u \stackrel{\leftarrow}{\thru D}
         \stackrel{\rightarrow}{D}^\lambda \sigma^{\rho\rho'} u \nonumber \\
         && + \bar u \sigma^{\rho\rho'}\stackrel{\leftarrow}{D}^\lambda 
       \stackrel{\rightarrow}{\thru D} u 
       + {3\over 4}\bar ugF^{\rho\rho'}\gamma^\lambda u
       + {3\over 4}  \bar dgF^{\rho\rho'}\gamma^\lambda d \ ,        
\end{eqnarray}
where $[\rho\rho']$ denotes antisymmetrization of the two indices.
The operator yields a contribution to $W^{\mu\nu\alpha}$,
\begin{equation}
      - {1 \over 12\pi^2 p^2}  \Pi_0(0,\mu^2) \ , 
\end{equation}
where $\Pi_0(q^2,\mu^2)$ is a two-point correlation function between 
$\hat O_5^{\lambda\rho\rho'}$ and ${\cal O}^\alpha$, 
and $\mu^2$ is an ultraviolet regulator to be
defined below.  

To calculate $\Pi_0(0,\mu^2)$, we again use the sum rule
approach. We first work out an operator-product expansion
for $\Pi_0(q^2)$ in the deep Euclidean space,
\begin{equation}
   \Pi_0(q^2,\mu^2)={\alpha_s \over 60\pi^3} q^4 \ln{\mu^2\over -q^2}
    + {1\over 12} \langle {\alpha_s\over \pi} F^2\rangle \ln{\mu^2\over -q^2}
    + {8\pi\alpha_s\over 9q^2} \langle \bar uu\rangle^2
      - {1\over 192\pi^2 q^2}\langle g^3G^3\rangle + ...
\end{equation}
On the other hand, we write a dispersion integral for $\Pi_0(q^2,\mu^2)$
valid for all $q^2$ \cite{bali2},
\begin{equation}
     \Pi_0(q^2,\mu^2) = {1\over \pi}\int^{\mu^2}_0 {ds\over s}
         {\rho(s)-\rho_{\rm pert}(s) \over s-q^2} \ . 
\end{equation}
where the upper limit defines the ultraviolet cut-off and 
\begin{equation}
        \rho_{\rm pert}(s) = {\alpha_s\over 60 \pi^2} s^3 \ .  
\end{equation}
$\Pi_0(q^2)$ defined in this way vanishes in perturbation theory and
its first power correction contributes in the same way as the last term 
in Eq. (\ref{dim4}). To find $\Pi_0(0,\mu^2)$, 
we assume a spectral function,
\begin{equation}
        \rho(s) = \pi f_R m_R^6 \delta(s-m_R^2) 
       + \theta(s-s_0) \left( {\alpha_s\over 60\pi^2} 
       s^3 + {\alpha_s\over 12}F^2 s \right)
\end{equation}
where $m_R$ is the mass scale for the exotic 
$1^{-+}$ resonace, suspected to lie between 
1.3 to 1.9 GeV \cite{exotica}. In our estimate, we take 
$m_R$ to be 1.5 MeV. The standard sum rule method allows 
us to extract $f_R$ = $1.8\times 10^{-3}$, which in turn
yields $\Pi_0(0,m_R^2) = 5.0 \times 10^{-4}m_R^2$.
The uncertainty of this number is at least a factor of 2 
due to unknown $m_R$ and the continuum threshold $s_0$,
which we take to be $1.9^2$ GeV$^2$.

\begin{figure}[htb]
\mbox{
\epsfxsize=16cm
\epsfysize=3cm
\hspace{0cm}
\epsffile{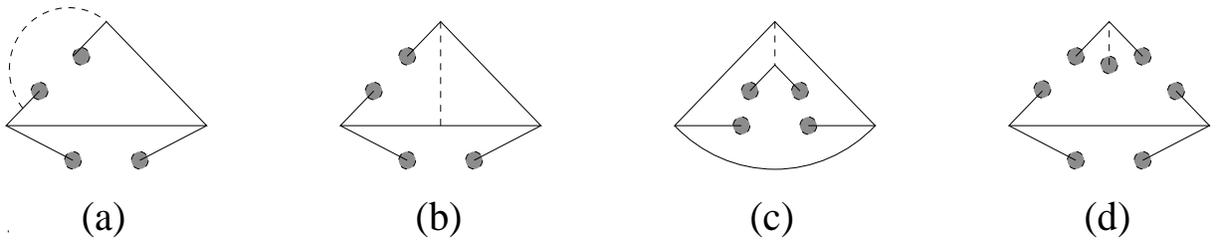}}
\vspace{0.3cm}
{\caption{\label{fig:fig3} Typical local (a,b,c) and bilocal (d) 
power corrections of dimension 6. }}
\end{figure}

The next term in the OPE for 
$W^{\mu\nu\alpha}$ involves dimension-six 
vacuum condensates, for which we use the 
factorization assumption. 
A calculation of the diagrams in Fig. \ref{fig:fig3}a,b,c (and similar ones 
which are not drawn)
give
a contribution to $W^{\mu\nu\alpha}$,
\begin{equation}
      {\alpha_s \langle \bar uu \rangle^2 
      \over 81\pi p^4}\left(20\ln{-p^2\over \mu^2}
               + 62 \ln{\mu^2\over -q^2}\right)\ , 
\end{equation}
where we have kept only logarithmic terms.
Small contribution of the first term 
to the final result justifies the approximation.  
The infrared logarithm in the second term 
signals that the contribution must be replaced by,
\begin{equation}
   {  4\langle\bar uu\rangle \over 3 p^4}\Pi_1(0,\mu^2)\ , 
\end{equation}
where $\Pi_1(q^2,\mu^2)$ is a bilocal correlator (see Fig \ref{fig:fig3}d)
give
involving $\hat {\cal O}^\alpha$ and the dimension-seven operator, 
\begin{equation}
        \hat  O_7^{\lambda\rho\rho'} = \epsilon^{ijk}\epsilon^{i'j'k}
            (D^\lambda u)^i C\gamma^\rho u^j ~\bar u^{j'}\gamma^{\rho'}
     C \bar u^{i'T} + {\it h.c.}
\end{equation}
The OPE for $\Pi_1(q^2,\mu^2)$ at large 
Euclidean $q^2$ is,
\begin{equation}
      \Pi_1(q^2) = {31\over 54}{\alpha_s\over \pi}\langle \bar uu\rangle 
       \ln {\mu^2\over -q^2} 
      - {m_0^2\langle \bar uu\rangle\over 3q^2} + ...
\label{pi1ope}
\end{equation}
where $m_0^2 = -\langle \bar u gF\cdot \sigma u\rangle/\langle
\bar uu\rangle$. The higher-order terms in ellipese 
involve condensates of dimension-seven and higher 
for which we know very little. To get an estimate, 
we assume vector-meson dominace \cite{bali1},
\begin{equation}
       \Pi_1(q^2) = {f_R' \over m^2_R - q^2} \ . 
\end{equation}
Expand the above in $q^2$ and matching its $1/q^2$ term with 
the OPE in Eq. (\ref{pi1ope}), we find,
\begin{equation}
       \Pi_1(0) = {m_0^2\langle \bar uu\rangle \over 3m_R^2}\ . 
\end{equation}

We ignore dimension-eight or higher contributions. 
In the factorization approximation, the 
contributions from dimension-eight condensates (both local and bilocal) 
are exactly zero. 

Based on the OPE we have developed for $W^{\mu\nu\alpha}$, 
we attempt an estimate for the $J_g(m_N^2)$. The sum rule 
equation reads like this,
\begin{eqnarray}
    {J_g \lambda^2 \over (m_N^2-p^2)^2} + ...
   & = &     {\alpha_s\over \pi^5}({1\over 144}
    \ln^2{-p^2\over \mu^2} - {1\over 36}\ln{-p^2\over \mu^2})p^2
          -{1\over 144\pi^2p^2} \langle {\alpha_s\over \pi}F^2\rangle
       \left(\ln{-p^2\over \mu^2} +{7\over 6}\right)
     \nonumber \\ &&
  - {1.1\times 10^{-3} \over 12\pi^2 p^2}
          + {20\alpha_s\over 81\pi} {\langle \bar uu \rangle^2 \over
p^4}\ln{-p^2 \over \mu^2}
      +  {4m_0^2 \langle \bar uu \rangle^2 \over 9m_R^2 p^4}
\end{eqnarray}
Substituting in the standard values for the condenstates 
at the normalization pont $\mu=1GeV$ (ccf ref. \cite{bbk} for example):
$\langle (\alpha_s/\pi)F^2\rangle = 0.012~ {\rm GeV}^4$, 
$\langle \bar uu\rangle = -0.017~{\rm  GeV}^3$, $m_0^2 = 0.65$, 
$\alpha_s(1 ~{\rm GeV}) = 0.37$, $32\pi^4\lambda_N^2 
= 2.5 ~{\rm GeV}^6$, $s_0 = 2.25 ~{\rm GeV}^2$, 
we find that the dimension-six bilocal
term is the dominant contribution. If we keep just 
this term, we find
\begin{equation}
       J_g(1~{\rm  GeV}^2) = {8e m_0^2 \langle \bar uu\rangle^2 \over 
       9m_R^2\lambda^2_N} = 0.25 \ . 
\end{equation}
A more careful analysis including other contributions yields, 
\begin{equation}
        J_g(1~{\rm  GeV}^2) = 0.35 \pm 0.13 \ . 
\end{equation}
where the error reflects the uncertainty of the mass 
scale in the $1^{-+}$ channel as well as
the uncertainty from the sum rule analysis. However, 
we have no way to know the accuracy of 
the vector meson approximation in estimating the dimension-six 
bilocal contribution.

The number we find, $J_g(1~{\rm GeV}^2) 
\sim 0.35\pm 0.13$  or $J_q(1~{\rm GeV}^2) \sim 0.15 \pm 0.13$, 
if taking seriously, has interesting implication on the spin structure of the 
nucleon. It says that gluons are at least as important in 
determining the nucleon spin as quarks, if not more.
Furthermore, from a recent globle analysis of data on polarized
deep-inelastic scattering\cite{altarelli}, one finds 
the gluon helicity $\Delta G(1~{\rm GeV}^2)$
defined in the infinite momentum frame and light-like gauge
has a size of 1 to 2 units of angular momentum. If correct, 
the gluon orbital contribution defined in a similar framework
must be large and negative and cancel a substential part 
of $\Delta G$. 
Such a large cancellation may be caused by the gauge-dependent
separation of $J_g$ into helicity and orbital contributions.
On the other hand, one half of the singlet-axial charge, or the quark spin 
contribution, is found to be $0.05^{+0.08}_{-0.05}$\cite{altarelli}. This leaves
about 20\% of the nucleon spin carried by quark orbital angular 
momentum. Here no large cancellation is present between the quark spin 
and orbital contributions.

\acknowledgements
This work is supported in 
part by funds provided by the
U.S.  Department of Energy (D.O.E.) 
under contracts
DOE-FG02-93ER-40762, DF-FC02-94-ER40818, and DE-AC05-84ER40150.

\end{document}